\documentclass[apj]{emulateapj}

\usepackage{color}

\usepackage{natbib,graphicx}

\usepackage{epsfig}
\usepackage{ulem}
\usepackage{color}
\usepackage{graphics,graphicx}
\usepackage{times} 
\usepackage{amssymb}
\usepackage{amsmath}
\usepackage{url}
\usepackage{multirow}
\usepackage{ctable}
\usepackage{threeparttable}
\newif\ifAMStwofonts
\AMStwofontstrue

\def\xmm{{\it XMM-Newton~\/}}

\def\swift{{\it SWIFT~\/}}
\def\integral{{\it INTEGRAL~\/}}

\def\epicpn{{\it EPIC}{\rm-pn}}
\def\epicmos1{{\it EPIC}{\rm-MOS1~\/}}
\def\epicmos2{{\it EPIC}{\rm-MOS2 ~\/}}
\def\epicmos{{\it EPIC}{\rm-MOS}}

\def\xmm{{\it XMM-Newton}}
\def\swift{{\it Swift}}

\def\integral{{\it INTEGRAL}}

\def\xspec{\hbox{\sc XSPEC}}

\def\s{\hbox{$\rm\thinspace s$}}
\def\ks{\hbox{$\rm\thinspace ks$}}
\def\ms{\hbox{$\rm\thinspace ms$}}

\def\hz{\hbox{$\rm\thinspace Hz$}}

\def\pc{\hbox{$\rm\thinspace pc$}}
\def\kpc{\hbox{$\rm\thinspace kpc$}}

\def\kmps{\hbox{$\rm\thinspace km~s^{-1}$}}
\def\pcmsq{\hbox{$\rm\thinspace cm^{-2}$}}

\def\ev{\hbox{$\rm\thinspace eV$}}
\def\kev{\hbox{$\rm\thinspace keV$}}

\def\ctsps{\hbox{$\rm\thinspace count~s^{-1}$}}

\def\ergpcmsqps{\hbox{$\rm\thinspace erg~cm^{-2}~s^{-1}$}}

\def\ergcmps{\hbox{$\rm\thinspace erg~cm~s^{-1}$}}

\def\msun{\hbox{$\rm\thinspace M_{\odot}$}}

\def\rg{${\it r}_{\rm g}$}
\def\rin{${\it r}_{\rm in}$}

\def\nh{${\it N}_{\rm H}$}

\def\chisq{{\chi^{2}}}

\def\laor{\rm{\small LAOR}}

\def\po{\rm{\small POWERLAW}}
\def\bb{\rm{\small BBODYRAD}}

\def\diskbb{\rm{\small DISKBB}}
\def\kerrbb{\rm{\small KERRBB}}

\def\comptt{\rm{\small compTT}}

\def\reflionx{\rm{\small REFLIONX}}
\def\refbhb{\rm{\small REFBHB}}
\def\tbnew{\rm{\small TBNEW}}

\def\kdblur{\rm{\small KDBLUR}}

\def\relconv{\rm{\small RELCONV}}
\def\relline{\rm{\small RELLINE}}

\def\xspec{\hbox{\small XSPEC~\/}}

\def\epchain{\hbox{\rm{\small EPCHAIN}}}

\def\epiclccorr{\hbox{\rm{\small EPICLCCORR}}}

\def\grid25{\hbox{\rm{\small GRID25}}}

\def\pile_est{\hbox{\rm{\small PILE_EST}}}

\def\epatplot{\hbox{\rm{\small EPATPLOT}}}

\def\j1118{\hbox{\rm XTE J1118+480}}
\def\j1749{\hbox{\rm J17497-2821}}
\def\jc{\hbox{\rm SWIFT~J1753.5-0127}}

\def\j1752{\hbox{\rm XTE~J1752--223}}

\def\cyg{\hbox{\rm Cygnus~X-1}}
\def\s{\hbox{\rm SWIFT~J1910.2--0546}}
\def\gx{\hbox{\rm GX~339-4}}

\begin{document}
\title[SWIFT~J1910.2--0546: retrograde spin or truncated disk?] {SWIFT~J1910.2--0546:  A possible black hole binary with a retrograde spin or truncated disk} 
\author{
R.~C.~Reis\altaffilmark{1,2},
M.~T.~Reynolds\altaffilmark{1}, 
J.~M.~Miller\altaffilmark{1},
D.~J.~Walton\altaffilmark{3},
D.~Maitra\altaffilmark{1},
A.~King\altaffilmark{1},
N.~Degenaar\altaffilmark{1}
}
\altaffiltext{1}{Dept. of Astronomy, University of Michigan, Ann Arbor, Michigan~48109, USA}
\altaffiltext{2}{Einstein Fellow}
\altaffiltext{3}{Space Radiation Laboratory, California Institute of Technology, Pasadena, CA 91125, USA}

\begin{abstract}

We present the first results from a long (51\ks) \xmm\ observation of the Galactic X-ray binary \s\ in a  intermediate state, obtained during its 2012 outburst. A clear, asymmetric iron emission line is observed and physically motivated models are used to fully describe the emission-line profile. Unlike other sources in their intermediate spectral states, the inner accretion disk in \s\  appears to be truncated, with an inner radius of \rin$=9.4^{+1.7}_{-1.3}$\rg\ at a 90\% confidence limit.  Quasi-periodic oscillations are also found at approximately 4.5 and 6\hz, which correlates well with the break frequency of the underlying broad-band noise. Assuming that the line emission traces the ISCO, as would generally be expected for an intermediate state, the current observation of \s\ may offer the best evidence for a possible retrograde stellar mass black hole with a spin parameter $a<{\rm -0.32 ~cJ/GM^2}$ ($90\%$ confidence).  Although this is an intriguing possibility, there are also a number of alternative scenarios which do not require a retrograde spin. For example, the inner accretion disk may be truncated at an unusually high luminosity in this case, potentially suffering frequent evaporation/condensation, or it could instead be persistently evacuated through mass loss in a relativistic jet. Further observations are required to distinguish between these different interpretations.

\end{abstract}


\section{Introduction}

From relativistic spectroscopy of the inner accretion disk \citep[e.g.][]{miller07review, Waltonreis2012}, to disk winds and jets \citep[e.g.][]{Miller2008j1655wind, miller2012cyg, Fender10, KingMiller2012, KingMiller2013}, to characteristic timescales \citep[e.g.][]{McHardy2006, agnqpo2008, Reis2012qpo}, black hole accretion appears to simply scale with mass. Indeed, some stellar-mass black holes are particularly good analogues of specific active galactic nuclei (AGN) classes.  By virtue of its high Eddington fraction and relativistic jets, GRS~1915$+$105 is considered to be the prototypical ``microquasar'', analogous to e.g.  3C 120 \citep{Chatterjee2009}. Similarly, V404 Cyg is a ``micro-LLAGN'' (Low-Luminosity AGN) somewhat akin to M81 \citep{Miller2010m81}. It is clear that through study of stellar mass black hole binaries, we can significantly increase our understanding of the most powerful engines in the universe. For recent reviews of the various spectral states observed in black hole binaries see \citet{Remillard06} and \citet{Bellonibook2010}.

Low-mass X-ray binaries are often observed to transit between different states, and the mass accretion rates onto stellar-mass black holes are seen to vary by  as much as $\sim 10^8$ between quiescence and outburst peaks \citep[e.g.][]{Dunn2010, Reynold2011swift}. During a typical outburst, these BHBs leave quiescence and pass through what has been termed  the low-hard state (LHS). Here the spectrum is seen to peak at $\sim100$\kev\ and consist mostly of a ($\Gamma\sim1.5$) powerlaw, with a  weak disk contributing less than 20\% of the total 2--20\kev\ flux \citep[see e.g.][]{Remillard06}. As the system progresses though the outburst, the luminosity increases and the spectrum softens ($\Gamma=1.4-2$). A hot ($\sim1\kev$) accretion disk often starts to dominate the 2--20\kev\ spectrum during the intermediate states (IS) and towards the peak of the outburst this component often contributes $>75\%$ of the total flux in what has traditionally been called the high-soft state (HSS). Although not all BHBs have been observed to  go through all the spectral states during their outburst, they often trace out a distinct pattern in a hardness-luminosity diagram (HLD; see \citealt{HomanStates2001, fenderetal04} for more details). 

The co-existence of a hard X-ray source, often referred to simply as ``the corona", a few \rg\ from an accretion disk \citep{reis2013corona} during the various active states, naturally leads to a further component in the energy spectrum as the hard X-rays are reprocessed in the relatively cool accretion disk \citep{rossfabian1993}. The strongest of these ``reflection" features are often associated with iron emission lines produced in the innermost regions of the  accretion disk. The profiles of these emission lines carry the imprint of the effects of strong Doppler shifts and gravitational redshifts natural to the space-time around black holes \citep{Fabian89,Fabian2000,laor}. A detailed study of reflection features allow us to place constraints on the distance between the line-emitting region and the black hole. 

In Boyer--Lindquist coordinates, the  radius of the  Innermost Stable Circular Orbit (ISCO) beyond which material can remain in a stable accretion disk around the black hole hole shrinks from $9GM/c^2$ for a maximal Kerr BH rotating in a retrograde sense relative to the accretion disk, i.e. having a dimensionless spin parameter $a=-1 cJ/GM^2$, to  $GM/c^2$ for a Kerr BH rotating in the prograde sense, with  $a=+1 cJ/GM^2$. In these units, the ISCO of a non-rotating, Schwarzschild black hole ($a=0$) is at $6GM/c^2$ \citep{Bardeenetal1972}. Thus it is clear that  studies of reflection features also  allow us to place constraints on  black hole spin, under the assumption that the emission region truncates at the ISCO \citep{reynoldsfabian08}. In this manner, ``spin" has been successfully measured for both stellar mass black holes \citep[e.g.][and references therein]{miller02j1650, miller07review, miller09spin} as well as for their supermassive counterparts \citep[e.g.][and references therein]{tanaka1995, FabZog09, 3783p1, Waltonreisspin2013,Risaliti2013Natur}.

\subsection{SWIFT~J1910.2-0546}

The new Galactic X-ray transient  \s\ was discovered by the \swift-BAT\ monitor on 2012 May 30 (MJD 56077; \citealt{ATel4347}; hereafter day 0). \swift\ follow up observations  starting on  June 1 2012 found the source to be very soft, with a disk fraction of 81\%, and suggested that \s\ belongs to the black hole binary family \citep{ATel4145, ATel4149}. Similar conclusions were presented by  \citet{ATel4198} where the authors find \s\ to be fully dominated by a soft, thermal disk component.  Optical spectroscopy \citep{ATel4210} found the system to  contain a featureless blue continuum and prompted the authors to compare \s\ with the short period black hole binary system  \jc\ \citep{Zurita2008j1753}.

As shown in Fig.~1, the source began a soft-to-hard state transition approximately on  2012 July 25 (day 56; \citealt{ATel4273}) and on August 3  we detected emission in radio (approximately 2.5~mJy at 6GHz; \citealt{ATel4295}; small arrow in Fig.~1) as it is usually seen in BHB systems during state transitions. During the $\approx45$ days that followed, as the system softened (see Fig.~1),  \s\ was also detected by \integral\ up to 200\kev\ (powerlaw model yields $\Gamma=1.8\pm0.3$; \citealt{ATel4328}) further establishing the compact object in \s\ as a black hole candidate. Nonetheless, it should be noted that at the time of writing there are no published dynamical mass estimates available for this source. By day 100, the system was back in a disk-dominated, soft state where it remained for approximately 30 days before it began to harden again. It was during this second soft-to-hard state transition that we obtained our \xmm\ observation, which is the subject of this paper. 

\section{Data Reduction}

We observed \s\ during the soft-to-hard state transition (Fig.~1) on 2012 October 17 (MJD 56217)  with \xmm\ for a total of 51\ks\ (ObsID  0691271401). The \epicpn\ camera \citep{XMM_PN} on-board \xmm\ was
operated in normal  ``timing'' mode with a ``thick'' optical blocking
filter and in the ``fast timing" submode with a frametime of 6\ms. In order to allocate full telemetry to the main instrument of interest, the observation was made with both  \epicmos\ cameras  turned off. Starting with the unscreened level 1 observation data files, we generated concatenated and calibrated event lists using the latest \xmm\ {Science Analysis System \thinspace v 12.0.1 (SAS)} and the tool \epchain. No background flares were observed during the observation, however the constant particle  background rate between 10--12\kev\ (PATTERN==0) was  higher than the optimum level\footnote{http://xmm.esac.esa.int/sas/current/documentation/threads/}  of 0.4\ctsps\ by approximately 50\%.

\begin{figure}
\label{figure1}
\begin{center}
{\hspace*{-0.7cm}
 \rotatebox{0}{
{\includegraphics[width=9.2cm]{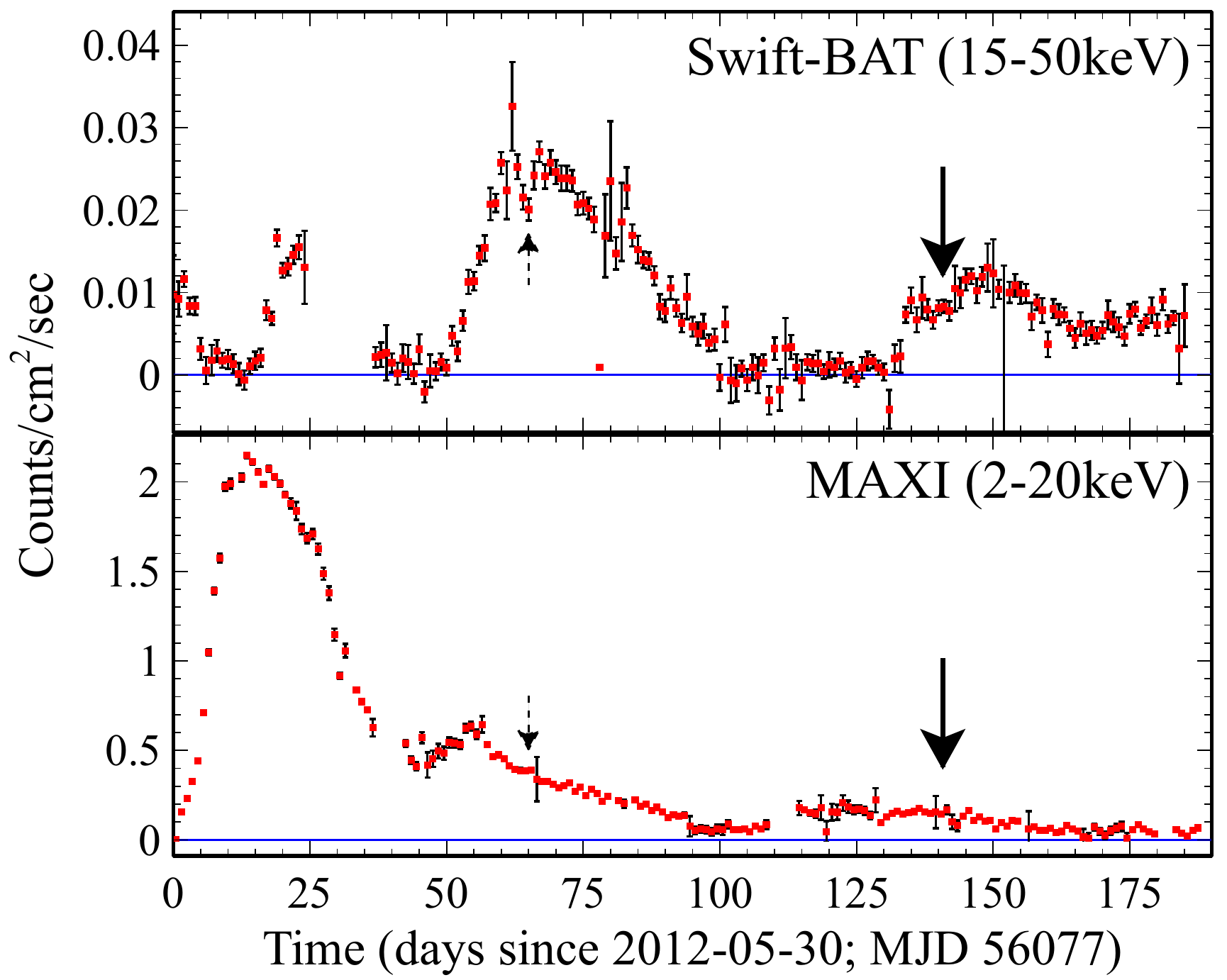} } 
}}
\caption{190--day light-curve of \s\ since its discovery on 2012-05-30 (day 0) in the 15-50\kev\ (top; \swift-BAT) and  2-20\kev\ (Bottom; MAXI). The large arrow shows the the time of the \xmm\ observation and the small arrow at $\approx65$ days shows the time of the radio detection. \swift\ and MAXI  lightcurves are available from: http://heasarc.gsfc.nasa.gov/docs/swift/results/transients/weak/SWIFTJ1910.2-0546/ and http://maxi.riken.jp/top/index.php?cid=1{\&}jname=J1910-057{\#}lsp respectively.  }
\end{center}
\end{figure}

{\color{black} In order to test the level of pile-up, we begin with a simple box region of width 18 pixels centred on the source and systematically excised a larger region from the centre of the PSF. For each region considered, we compared the pattern distribution with the  SAS task \epatplot\ and found that  an exclusion region of 5 pixels resulted in a maximum pile-up level of 2\%. However, because the effects of pile-up depend on the shape of the incident continuum \citep{pileup2010}, in addition to comparing the pattern distribution, we also directly compare the spectra extracted with each of the regions considered.}

{\color{black} All spectra were extracted after excluding bad pixels and pixels at the edge of the detector, and we only consider single and double patterned events. Response files were created in the standard manner using RMFGEN and ARFGEN. Because of the high source flux in the EPIC-pn spectrum we do not subtract any background when fitting the energy spectrum. Finally we rebinned the spectrum with the tool GRPPHA to have at least 25 counts per channel. The spectra extracted from the five different regions are shown in Fig. 2. To compare the spectra, we apply the same absorbed powerlaw plus disk blackbody model to each, only allowing a normalisation constant to vary between them. In fitting the continuum, the 4--8\kev\ range was ignored in a similar manner to that presented for \gx\ in \citet{reisgx}. The  effects of pile-up can be clearly seen in that the breadth of the iron line decreases as the level of pileup increases. This artificial ``narrowing" of the line profile  is exactly as predicted for X-ray binaries  in \citet{pileup2010}.} We also investigated the effect of allowing the continua to vary fully by also allowing the photon index and disk temperature to vary between the different spectra. As expected, the piled-up spectrum with no inner pixels excluded is found to be harder ($\Gamma_{\rm 0px}=2.173\pm0.007$) than that with either 1, 2, 5 or 7 pixels removed  ($\Gamma_{\rm 1px}=2.192\pm0.008$; $\Gamma_{\rm 2px}=2.23\pm0.01$; $\Gamma_{\rm 5px}=2.241\pm0.015$; $\Gamma_{\rm 7px}=2.24\pm0.018$), although the evolution in $\Gamma$ is not very extreme in this case. Upon noticing the iron line region, the results described above are again observed; the breadth of the iron line decreases as the level of pileup increases.

These results fully support the conclusion obtained through comparison of the pattern distributions, in that excising a region of 5 pixels is sufficient to remove the effects of pile-up, as the results obtained are consistent with the more conservative extraction region of 7 pixels, suggesting the pile-up has indeed been minimised.  In order to maximise signal-to-noise we proceed with the 5~pixels exclusion region, but note that our results remain unchanged if we adopt the 7~px region.

\begin{figure}
\label{figure2}
\begin{center}
{\hspace*{-0.5cm}
 \rotatebox{270}{
{\includegraphics[width=6.7cm]{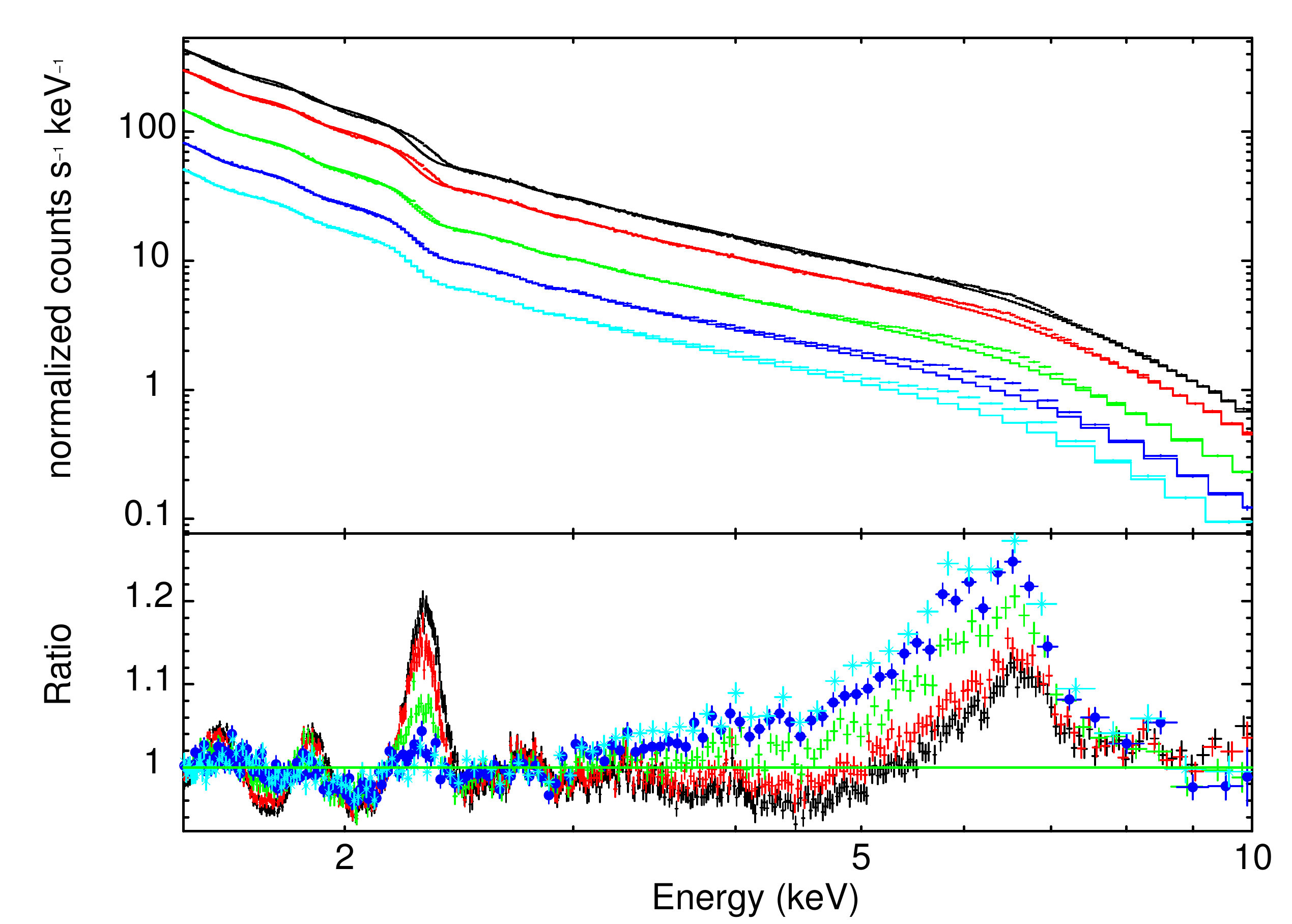} } 
}}
\vspace{-0.2cm}
\caption{Spectra for various source extraction regions fitted with an absorbed powerlaw plus disk black-body model. (Top) From top-to-bottom the spectra are for: full 18~px region centred on the source (black); single central pixel excised (red); central 2 pixels excised (green); central 5~pixels excised (blue); central 7~pixels excised (cyan). The energy range 4-8\kev\  was ignored during the fit. The narrowing of the emission line profile caused by excessive pile up is clear. Pile up becomes insignificant beyond the central 5~pixels (blue and cyan). Note also that the instrumental line at $\approx2.2$\kev\ also decreases significantly as the level of pile up decreases.  }
\end{center}
\end{figure}

Source and background\footnote{In making the lightcurves, we used background regions centred at RAWX=4 having a width of 2~pixels as suggested in the \xmm\ Calibration Technical Note for \epicpn\ fast modes {http://xmm2.esac.esa.int/docs/documents/CAL-TN-0083.pdf}. } light curves were created  with a time resolution of $6\times10^{-3}$~s and all the necessary corrections were applied using the task \epiclccorr\ in accordance with the \xmm\  data reduction guide\footnote{http://xmm.esac.esa.int/sas/current/documentation/threads/timing.shtml}. We note here that the timing results are not changed upon the exclusion of the background lightcurves. Throughout this paper, the  \epicpn\ spectrum is fit in the 1.5--10.0\kev\ energy range. This lower limit was chosen as it was found that the data below this energy showed strong positive residuals above any reasonable continuum. Similar residuals were reported by \citet{hiemstra1652} for XTE~J1652-453 and \citet{reis1752} for \j1752, both times based on data obtained with  \epicpn-timing\ mode. This ``soft excess" was suggested by these authors to be  associated  with possible calibration issues related with the redistribution matrix of the timing-mode data, but its true origin remains unknown\footnote{See http://xmm.vilspa.esa.es/docs/documents/CAL-TN-0083.pdf}. The energy range 2.1--2.4\kev\ was also ignored as it contained the obvious presence of a feature associated with instrumental Au edges (Fig.~2). All errors reported in this work are 90~per~cent confidence errors obtained for one parameter of interest unless otherwise noted.

\section{Data Analyses and Results}

\begin{figure}[!t]
\label{figure3}
\begin{center}
{\hspace*{-0.5cm}
 \rotatebox{0}{
{\includegraphics[width=8.2cm]{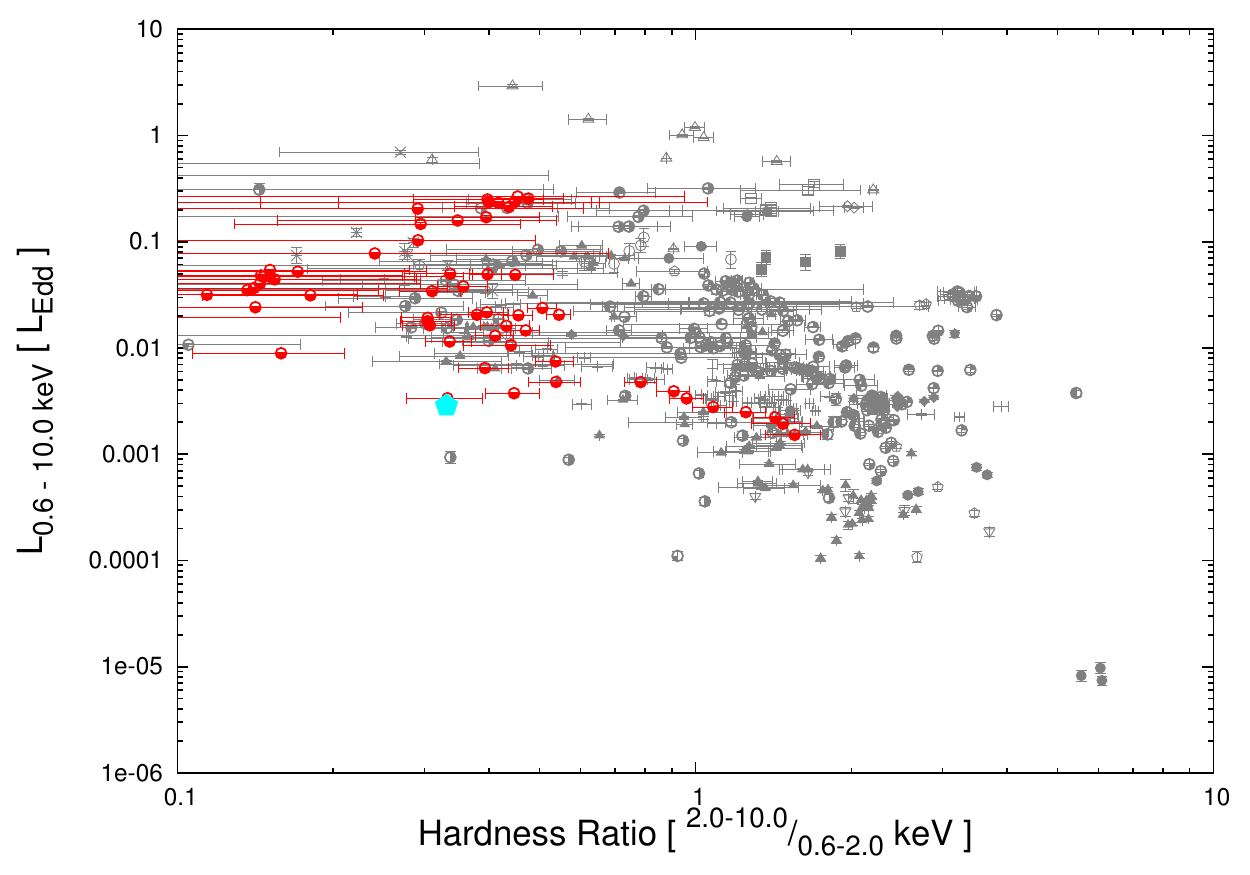} } 
}}
{\hspace*{-0.5cm}
 \rotatebox{0}{
{\includegraphics[width=8.2cm]{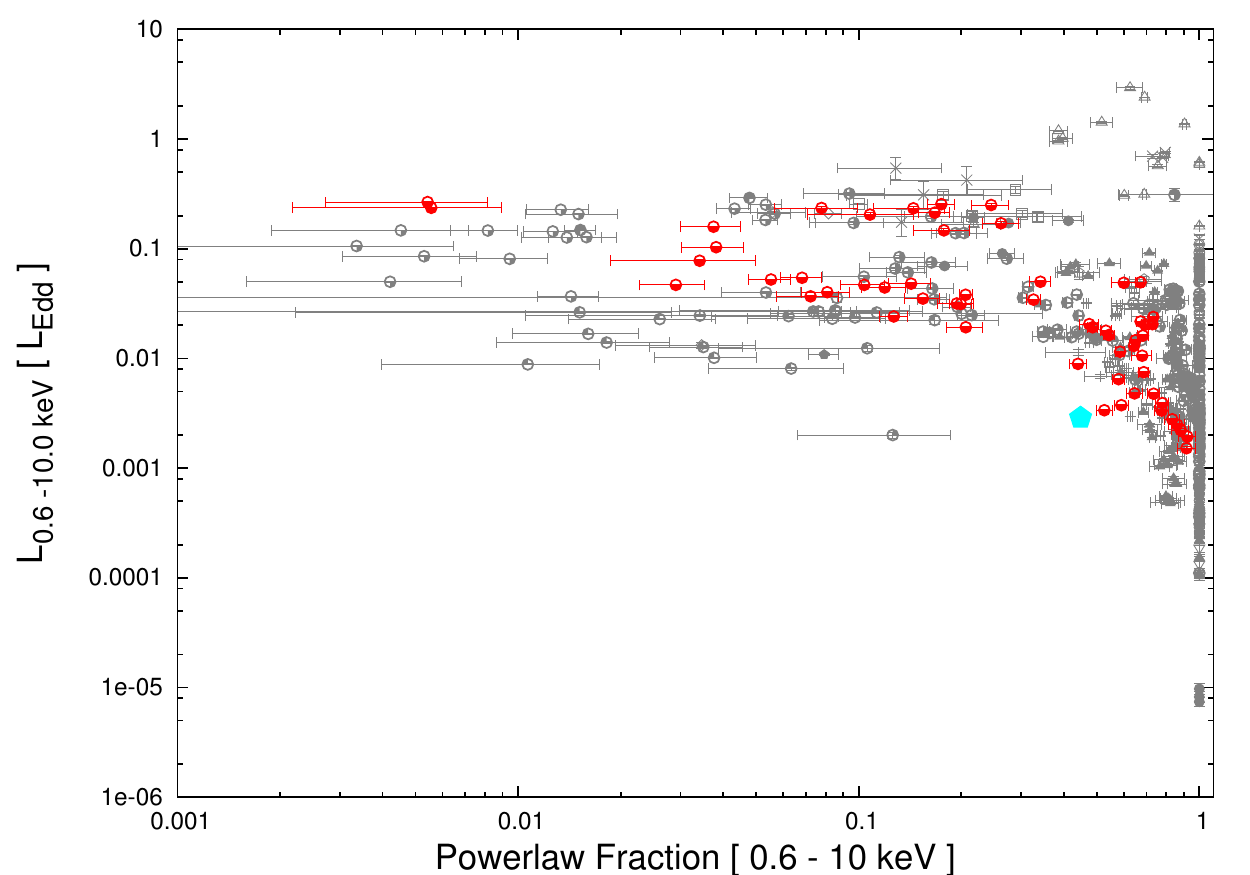} } 
}}
\caption{Evolution of \s\ (red) as seen with \swift\ and expressed in terms of the hardness luminosity diagram (top) and disk fraction luminosity diagram (bottom). In both cases, it is clear that \s\ behaves in a similar manner to all other confirmed and candidate black hole binaries (grey sample from \citealt{Reynold2011swift}).  The \xmm\ observation of \s\ are shown in cyan for comparison. }
\end{center}
\vspace{-0.5cm}
\end{figure}

In order to trace out the complex evolution of the outburst, we show in Fig.~3 (top) the Hardness-Luminosity Diagram \citep{HomanStates2001, fenderetal04}. To make this diagram, we used  \swift-XRT\ pointing observations (Target ID 32521) which monitored the system over its outburst every $\approx 3$~days for a total exposure of $\approx 50$\ks. The hardness ratio is calculated from the ratio of the unabsorbed hard to soft X-ray fluxes, obtained using a simple model consisting of an absorbed disk plus power-law. In scaling the 0.6--10\kev\ luminosity to the Eddington value, we have assumed a black hole mass of 8\msun\ as per the peak of the black hole mass distribution found in \citet{ozelbhmass2010} and \citet{ Kreidbergbhmass2012} together with a distance of 6\kpc. These values are broadly consistent with the independent estimates made later in the paper (see \S~4 and Fig.~11).  \s\ -- shown in red --  traces out the typical ``q" shaped pattern observed for various black hole binaries, as exemplified by the grey points from the \swift\ sample of  stellar mass black hole candidates \citep{Reynold2011swift}.

The disk fraction luminosity diagram \citep[see e.g.][]{Kalemci2004, Kalemci2006, Tomsick2005, Dunn2010, Reynold2011swift}  shown in Fig.~3 (bottom) also confirms that the system enter phases of extreme softness, characteristic of BHB transients. We defer a detailed analyses of the outburst as seen with \swift\ to the near-future and concentrate here on the deep \xmm\  observation. Below we present the results of our timing and spectral analyses from this data.

\subsection{Power Density Spectrum}

Starting from the total \epicpn\ time series with a time resolution of $0.012$~seconds in the full 0.2-12\kev\ energy range, we sub-divided the data into lengths of 512 seconds long and created a power density spectrum (PDS) for each segment, normalized after \citet{Leahy1983}. This was then averaged to produce a single power spectrum which was geometrically re-binned and  had the Poisson  level  subtracted. Visual inspection immediately showed the presence of band-limited noise which could be adequately modelled with a twice-broken power-law together with a broad bump between $\approx 2-7$\hz. Limiting the data to energies above 1~\kev\ enhances this bump and a possible QPO appears at around 4--6\hz\ (Fig.~4) . The band-limited noise can be adequately fit in the full frequency range considered here (0.16--41.7\hz) with a twice-broken  power-law,  once the 2--7\hz\ region is removed. This model ($\chisq/\nu$=115.2/107) follows a flat powerlaw below a break frequency of  $\nu_{b}=1.05\pm0.27$\hz, at which point the index is roughly 1 until the second break around 4\hz\ where the index steepens to $>2$.

\begin{figure}
\label{figure4}
\begin{center}
{\hspace*{-0.5cm}
 \rotatebox{0}{
{\includegraphics[width=8.cm]{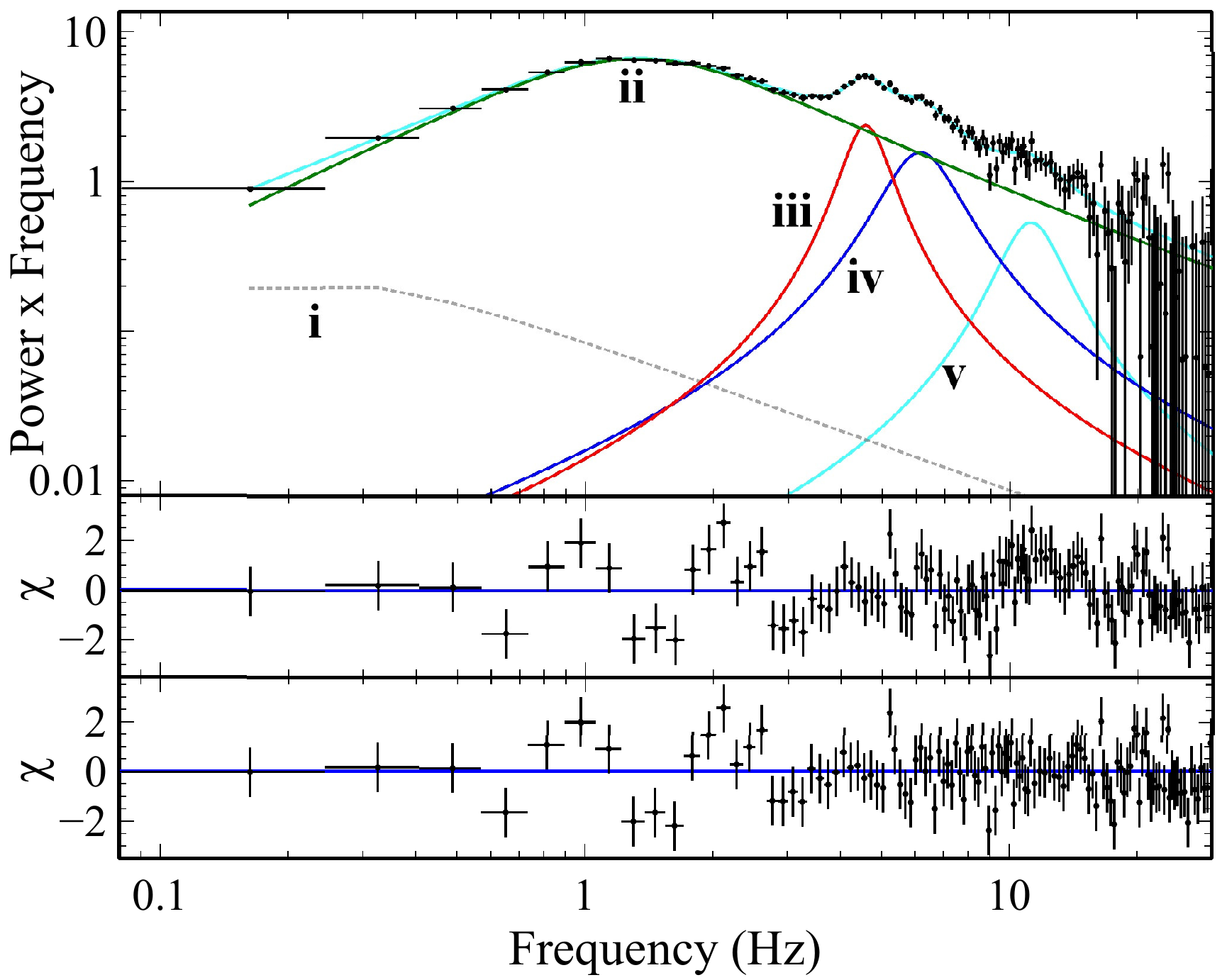} }}}
\caption{Power Density  Spectrum of \s\ in the 1--12\kev\ band fit with 5 Lorentzian (top). The middle and bottom panels show the residuals to the best fit in units of $\sigma$  for a model with 4 and 5 Lorentzian, respectively. In the middle panel, the least significant $\sim11$\hz\ Lorentzian (v) was removed.}
\end{center}
\end{figure}

Following \citet{Nowak2000qpo}, we proceed by modelling the power spectrum of \s\ as a sum of Lorentzian components, and show in Fig.~4 the fit to the full frequency range ($\chisq/\nu$=150.3/130) with 5 Lorentzian components (i--v), with the first two used to model the band-limited noise described above and the latter 3 describing the QPO complex around 6\hz\ in a similar manner to \citet{Bellonibook2010}. We emphasise here that this model now represents the  full  0.16--41.7\,\hz band as opposed to the previous twice-broken  power-law model which ignored the  2--7\,\hz\ region. The centroid frequencies (widths) of the various Lorentzians are:\footnote{The corresponding values for the power spectrum produced without background are: (i) $0 (<0.7)$\hz, (ii) $0.75^{+0.05}_{-0.04} (2.19\pm0.05)$\hz, (iii) $4.53\pm0.05 (1.1\pm0.2)$\hz, (iv) $6.0^{+0.3}_{-0.2} (2.5^{+0.6}_{-0.5})$\hz, and (v) $11.0^{+0.6}_{-0.7} (4.3^{+1.9}_{-1.4})$\hz, in excellent agreement with the values presented in the main text.}  $0 (<0.7)$\hz, (ii) $0.75\pm0.05 (2.18\pm0.05)$\hz, (iii) $4.53\pm0.07 (1.2\pm0.3)$\hz, (iv) $6.0^{+0.3}_{-0.4} (2.6\pm0.7)$\hz, and (v) $11.0^{+0.7}_{-0.8} (3.9^{+2.4}_{-1.6})$\hz.

 {\color{black}Interestingly, the first two relatively coherent features at $\sim4.5$ and $\sim6$\hz\ follow roughly the 3:2 ratio often observed at higher frequencies in black hole binary QPOs \citep[e.g.][]{Abramowicz200132}, however the last feature at $\sim11$\hz\ does not fall in any obvious harmonic relation. Nevertheless, removing this Lorentzian worsens the fit by $\Delta\chisq/\Delta\nu=30.5/3$.}

Figure~5 shows that the results obtained above for \s\ follow the same trend as all black hole candidates and  low-luminosity neutron star systems (\citealt{qpofvbreak1999}; see also \citealt{Belloni2002apobreak}) in that the frequency of the low-frequency QPO correlates strongly with break frequency\footnote{In making this qualitative figure, we used the Dexter Java application of \citet{dexter_ads} to estimate the values of QPO and break frequencies from  Figure~3 of \citet{qpofvbreak1999} and superimposed our break frequency, as found from the  twice-broken  power-law model described above ($\nu_{b}=1.05\pm0.27$\hz), together with the first potential harmonic at $4.53\pm0.07$\hz.}.

\subsection{Energy Spectrum}

All evidence presented so far points towards \s\ being a stellar mass black hole candidate which has been caught by \xmm\ during a transition from the soft to the low-hard spectral state. For this reason, we start by modelling the energy spectrum with a combination of an absorbed powerlaw together with a disk black-body component. For the neutral absorption, we use the model \tbnew\footnote{http://pulsar.sternwarte.uni-erlangen.de/wilms/research/tbabs/} together with the ISM abundances of \citet{Wilmsabun} and cross sections of \citet{Vernerxsec}. Visual inspection of the residuals showed the clear presence of an excess to the continuum around 6\kev\ which we interpret as fluorescent iron emission, as is often seen in  accreting black hole binary systems (e.g. \citealt{DiazTrigo2007j1655, hiemstra1652,  Fabian2012cyg}; for a recent review see \citealt{miller07review}).

\begin{figure}[!t]
\label{figure5}
\begin{center}
{\hspace*{-0.5cm}
 \rotatebox{0}{
{\includegraphics[width=8.cm]{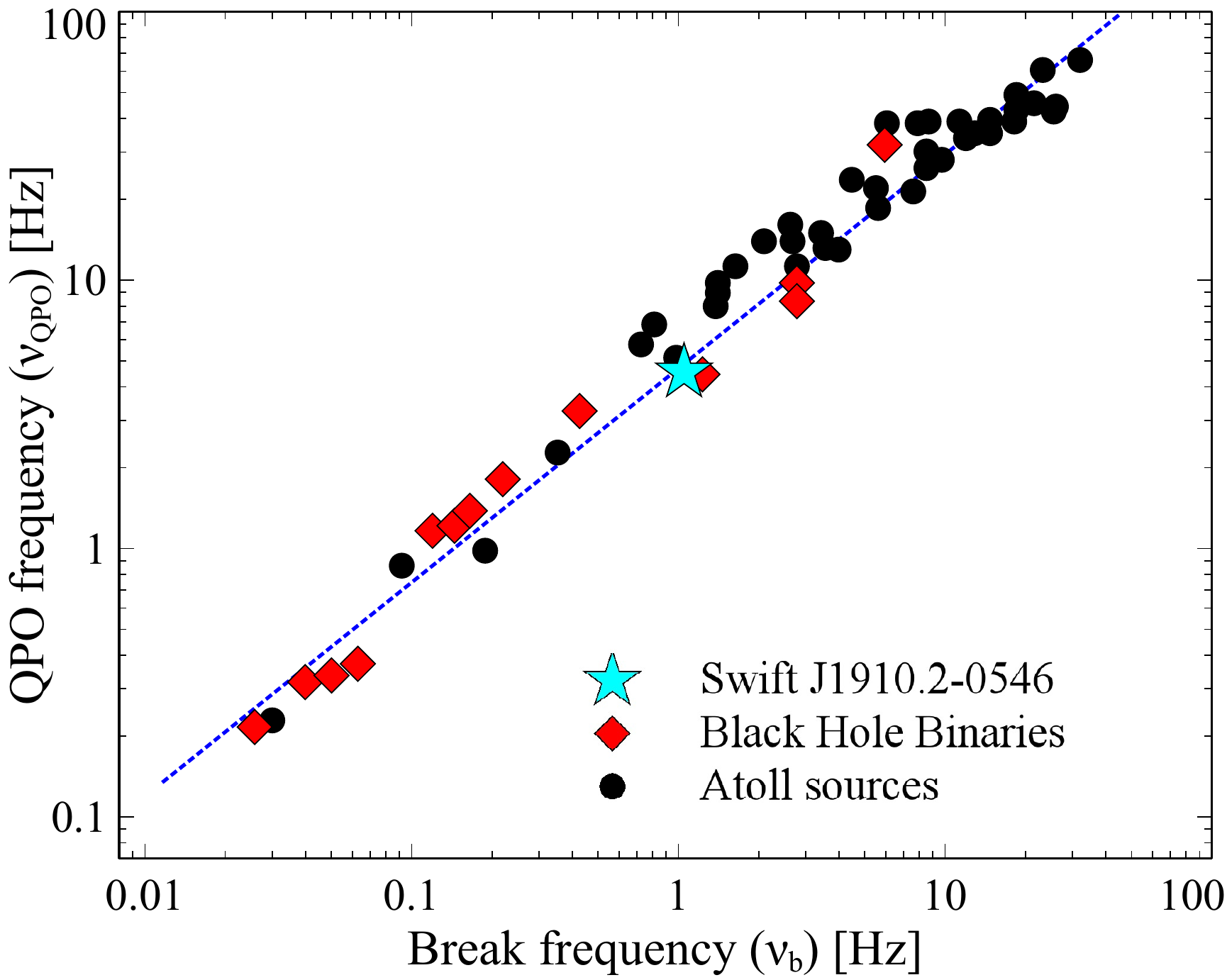} }}}
\caption{Break versus QPO frequency for a sample of BHBs (red) and NS-Atoll sources (black) after \citet{qpofvbreak1999}, with \s\ shown in cyan for comparison. (See footnote 7) }
\end{center}
\end{figure}

\subsubsection*{Phenomenological view of the iron line}

In Fig.~6, we show the data/model ratio to a number of different continuum after excluding the 4--7\kev\ energy range. The models shown are  (i) \po\ + \diskbb\  (\diskbb\ model of \citealt{diskbb}); (ii) \po\  + \bb\ (blackbody spectrum); (iii) \diskbb\ + \bb\ + \bb; (iv) \comptt + \diskbb\ (\comptt\ model of \citealt{comptt}). It is clear that the profile of this feature remains essentially unchanged independent of the chosen continuum.  It  is also apparent that the excess is asymmetric (which remains true when plotted on a linear energy scale). 

In order to further demonstrate this, we begin by adding a Gaussian profile to model the excess. In doing so, we initially limit the centroid energy of the line to 6.4--6.97\kev, encompassing the full range of possible ionisation states of iron.  The data-to-model ratio  to this fit  ($\chisq/\nu=1805.5/1607 =1.12$) is shown in the middle panel of Fig.~7. As expected, there remains obvious residuals between 5--7\kev\ as the symmetric Gaussian profile -- whose centroid energy pegs at the lower energy limit of 6.4\kev -- can not account for the excess red-wards of the peak, despite being broad ($\sigma = 728 \pm 65\ev)$. Allowing the energy to be less than 6.4\kev\  results in an un-physical line energy ($6.11\pm0.04$\kev) and again requires an extremely broad profile with $\sigma = 770 \pm 60\ev$.

\begin{figure}[!t]
\label{figure6}
\begin{center}
{\hspace*{-0.5cm}
 \rotatebox{0}{
{\includegraphics[width=8cm]{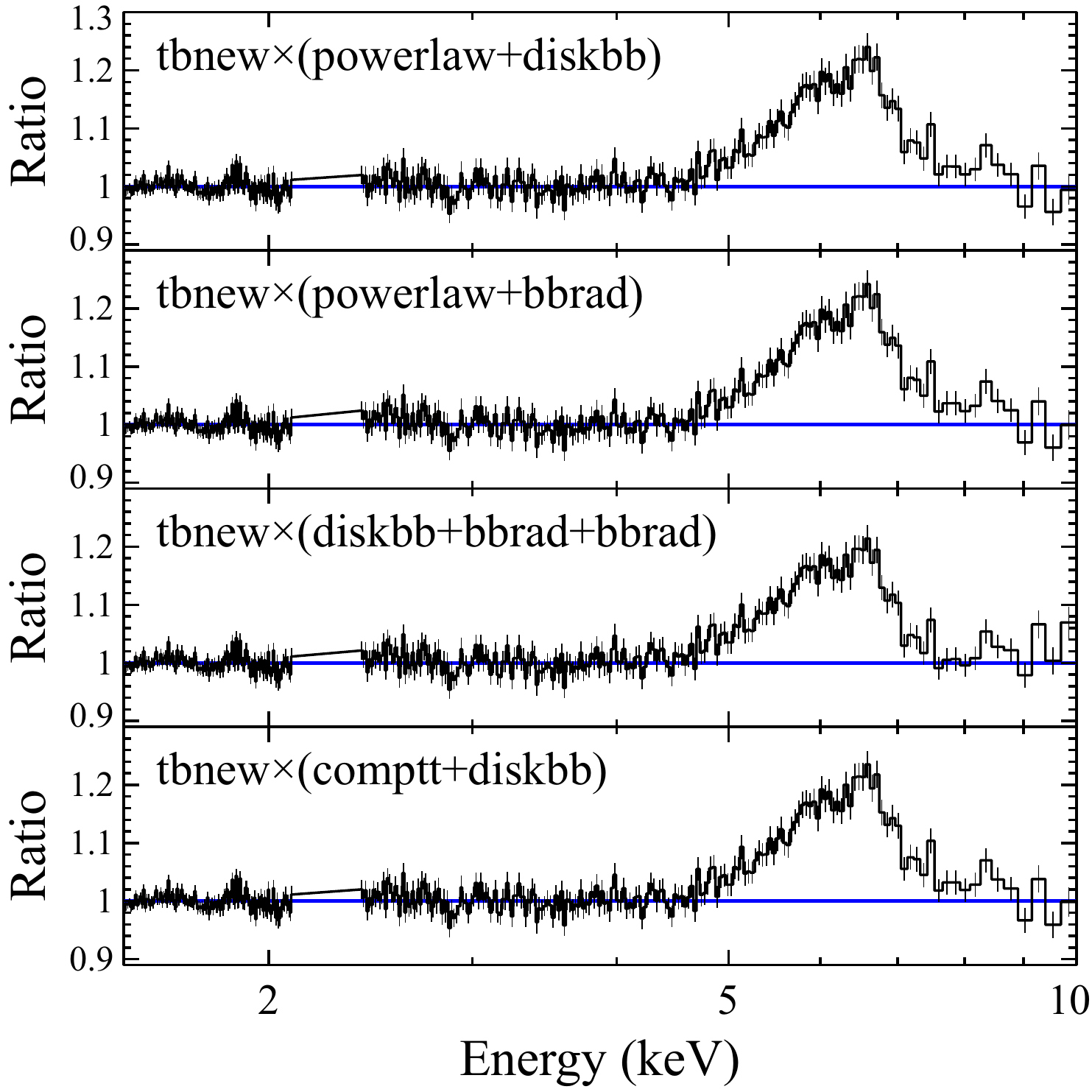} }}} 
\caption{Data/model ratio to a variety of absorbed Compton components together with thermal disk emission. The various models were fitted in the 1.5--4.0\kev\ and 7-10\kev\ energy range. A broad, asymmetric  iron-emission line extending to approximately 5\kev\ is present  independently of the model used for the hard emission. The data
have been rebinned for plotting purposes. }
\end{center}
\end{figure}

It has been shown by numerous authors that such a broad line profile is not likely to be solely broadened by Compton scattering (e.g. \citealt{fabianetal95,reynoldsandwilms2000, milleretal2004Achandra}, see also \citealt{hiemstra1652,Waltonreis2012}). We therefore proceed by assuming a relativistic origin for the broad emission line. We replace the Gaussian line  with a \laor\ model \citep{laor} which describes the line profile  expected from the inner disk around a maximally spinning black hole. The line energy is again  constrained to lie between 6.4 and 6.97\kev,  and we begin by allowing the emissivity profile of the model to follow a powerlaw  profile such that $\epsilon(r)\propto r^{-q}$ where $q$ is free to be any positive value. The outer disk radius was frozen at the maximum value in the model of 400\rg. This model, detailed in Table~1 (Model~1a) and shown in Fig.~7 (top and bottom panels), resulted in a satisfactory fit with ($\chisq/\nu=1703.51/1605=1.06$), although there are still some minor residuals in the iron K bandpass. 

The inner-disk radius found here of $r_{\rm in} = 5.3^{+0.2}_{-0.4}$\rg, if associated with the radius of the ISCO, is suggestive of a black hole that is not spinning rapidly (but see \S4 for other interpretations), and the near-Newtonian value for the emissivity index ($q=3.5\pm0.2$), is consistent with this hypothesis \citep[e.g.][]{wilkins2011}\footnote{The emissivity profile under normal Newtonian approximations at $r\gg r_{\rm g}$,  is expected to be flat in the region directly  below the illuminating, isotropic source (the corona), while tending to $r^{-3}$  when $r>>h$, where the flux received by the disc from the source falls off as the inverse square of the distance with a further factor of $1/r$ arising from the cosine of the angle projecting the ray normal to the disc plane. See \citet{wilkins2011} for more details on the expected emissivity profiles around black holes.}. In Table~1, we also detail the results from this model when the emissivity index is frozen at the Newtonian value of 3 (Model~1b), as will be assumed for much of the remainder of this work. In all models considered so far, the \diskbb\ component consistently required a temperature of approximately 0.3\kev. The total 2--20\kev\ unabsorbed flux is $2.2\times 10^{-10}\ergpcmsqps$ of which approximately  8~per~cent is associated with the accretion disk.

Before moving on to model the line profile in a self-consistent manner, we stress  that we have first made use of the well known \laor\ model in order to facilitate straightforward comparison with previous, published  literature,  but also due to the standardised output parameters. Nonetheless, when replacing this model with the more sophisticated \relline\ profile (\citealt[][]{relconv}), we obtain identical values to those quoted in Table~1. Specifically, when freezing the spin of the black hole -- a free parameter in \relline\ -- to its maximal prograde value, and allowing the inner accretion disk radius to vary, we find; $r_{\rm in} = 5.5^{+0.3}_{-0.4}$\rg, $\theta =17\pm1$~degrees, and all other parameters are similarly consistent with those of the \laor\ model.

\begin{figure}[!t]
\label{figure7}
\begin{center}
{\hspace{-0.5cm}
\rotatebox{0}{
{\includegraphics[width=8.cm]{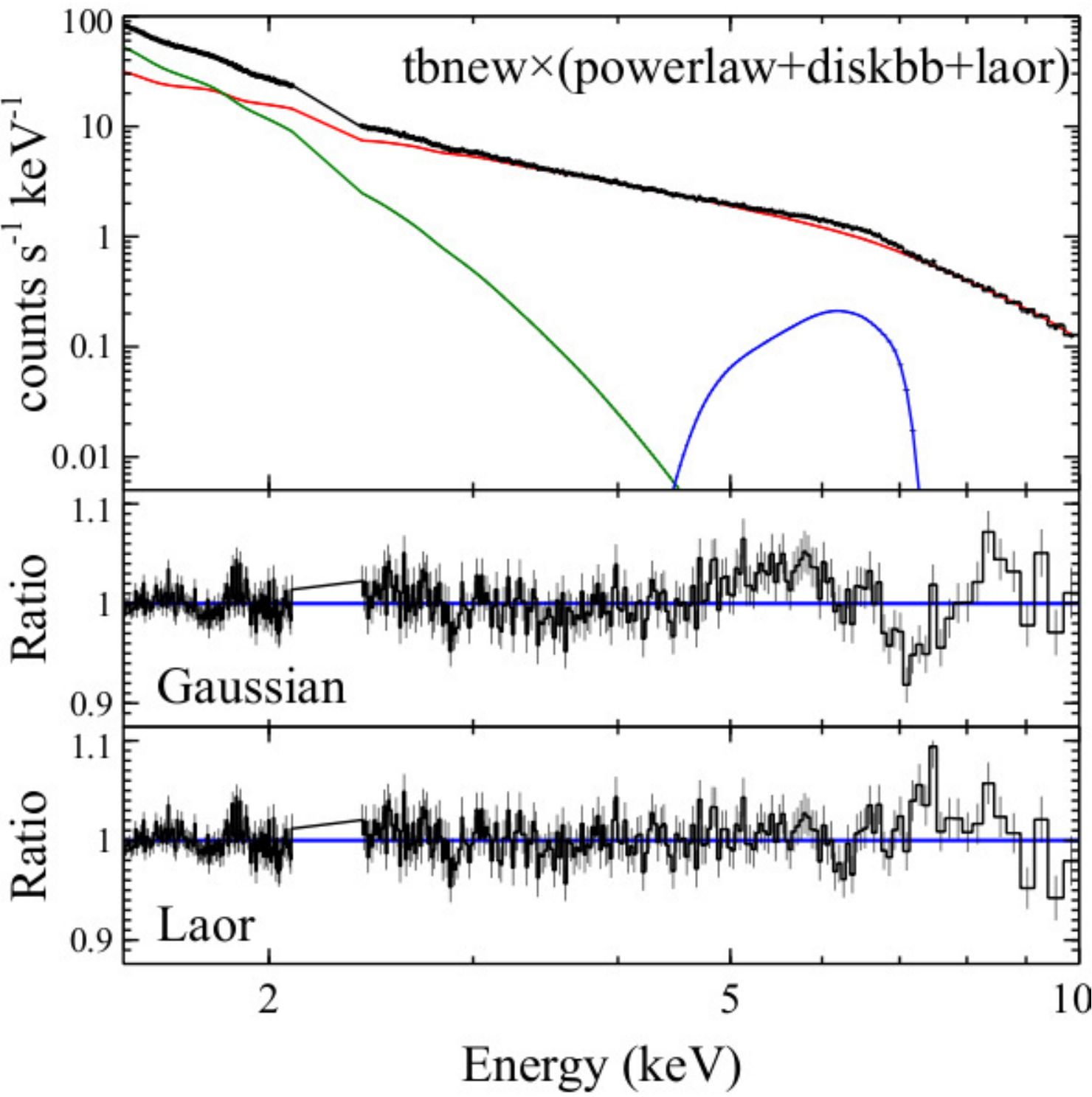}}}}
\caption{(Top:) Spectrum of \s\  fit with an absorbed power-law (red) together with a disk component (green) and a relativistic emission line (blue; Model 1a). (Middle:)  Ratio of the data to a model consisting of Gaussian profile to model the emission line. (Bottom:)  Data/model ratio to the model shown in the top panel (Model 1a). It is clear that the symmetric Gaussian profile is a poor description to the excess observed around 6\kev. The spectra have been binned
for visual clarity only.}
\end{center}
\end{figure}

\begin{table*}
\begin{center}
\caption{ Summary of model parameters}
\label{table2}
\begin{tabular}{lcccccccc}                
  \hline
  \hline 
  
& Model~1a  & Model~1b & Model~2  & Model~3a & Model~3b & Model~3c  \\

\nh\ (~$\times10^{22}$\pcmsq) & $0.26\pm0.07$ & $0.28^{+0.06}_{-0.08}$ &  $0.26^{+0.08}_{-0.07}$ &$0.15^{+0.18}_{-0.12}$&$0.19^{+0.17}_{-0.16}$ & $0.15^{+0.06}_{-0.12}$ \\

$\Gamma$ 									&$2.236\pm0.015$ & $2.229^{+0.011}_{-0.008}$ &$2.21\pm0.02$ & $2.14\pm0.02$& $2.14^{+0.03}_{-0.02}$ & $2.14\pm0.01$ \\
$N_{\rm powerlaw}$ 					& $0.081\pm0.002$   & $0.080^{+0.001}_{-0.002}$ & $0.061^{+0.004}_{-0.006}$ &$0.051^{+0.005}_{-0.003}$&$0.051^{+0.005}_{-0.004}$ & $0.051\pm0.002$ \\

$kT_{\rm disk}$(\kev)  			& $0.302^{+0.006}_{-0.005}$ & $0.301^{+0.006}_{-0.004}$ & $0.300\pm0.006$ & $0.217^{+0.004}_{-0.005} $& $0.219\pm0.003$ & $0.218^{+0.001}_{-0.004}$ \\
$N_{\rm diskbb}$ ($\times10^3$)& $4.5^{+0.8}_{-0.7}$&$4.7^{+0.7}_{-0.8}$& $4.6^{+1.0}_{-0.7}$ & --- & --- & --- \\

$E_{\rm Laor}$(\kev)  					& $6.97_{0.05}$  &  $6.81\pm0.04$ & --- & ---& --- & --- \\
$N_{\rm Laor}$  ($\times10^3$)	& $0.43\pm0.03$  &  $0.39^{+0.03}_{-0.02}$ & --- & ---& --- & --- \\

$N_{reflionx}$ ($\times10^{-7}$) & --- & ---  &$1.8^{+1.1}_{-0.4} $ & ---  & --- & --- \\
$\xi$ ($\ergcmps$     )$^a$											& --- & --- &$2100\pm700$ & $4290\pm920$ &$4450\pm580$ & $4840\pm670$ \\

$F_{Illum}/F_{BB}$ 					    & ---& ---&---& $0.8^{+0.1}_{-0.2}$& $0.8\pm0.2$ & $0.83\pm0.09$ \\
$n_{H}$ ($\times10^{19}$)      &--- & --- &--- & $1.2^{+0.2}_{-0.3}$& $1.2^{+0.2}_{-0.1}$ & $1.1^{+0.2}_{-0.1}$ \\
$N_{refbhb}$                               & ---  & --- & ---  &$0.027\pm0.002 $&$0.027\pm0.002$ & $0.027\pm0.002$ \\

$q$ 									   			& $ 3.5\pm0.2$ & 3(f) &3(f) & 3(f)& 3(f) & $2.7\pm0.1$ \\
$\theta$ (degrees) 						& $16^{+2}_{-6}$ & $9^{+9}_{-5}$ & $19^{+3}_{-4}$ & $17\pm4$& $20^{+2}_{-3}$ & $7^{+4}_{-3}$ \\
\rin\ (\rg) 									&$5.3^{+0.2}_{-0.4}$ & $5.0^{+0.5}_{-0.2}$&$8^{+2}_{-1}$&$9.4^{+1.7}_{-1.3}$& --- & --- \\
Spin ($a$) 									&---& --- &--- & ---& $<$ $-0.82$ & $<$ $-0.32$\\
$\chi^{2}/\nu$                          &1703.5/1605   & 1744.5/1606 & 1664.9/1606 & 1655.9/1606&1656.6/1606 & 1647.9/1605 \\
\hline
\hline
\end{tabular}
\end{center} 
\small Notes: $^a$The ionisation parameters for Model~3 are not from direct fitting and were estimated as described in the text. Model~1a is described in \xspec\ as $\tbnew*(\diskbb + \po +\laor)$. Model~1b is identical to 1a but for the emissivity index frozen at 3. Model~2 replaces the \laor\ profile  with the reflection model \reflionx\ convolved with \kdblur. In \xspec it is described as $\tbnew*(\diskbb + \po +\kdblur*\reflionx)$. Model 3a now replaces both \diskbb\ and \reflionx\ with  reflection model \refbhb\ which incorporates both. In all models described so far the kernel from the \laor\ line profile was used to account for the gravitational effects close to the black hole. In Model~3b, we replace the \kdblur\ kernel with the  relativistic code \relconv\ were the spin is a parameter of the model and allowed to be negative (retrograde spin). Model~3c now allows the emissivity index to be different than the Newtonian value of 3.  In all cases the hard emission illuminating the disk is assumed to be a powerlaw with index $\Gamma$. All errors are 90~per~cent confidence for one parameter of interest. 
\end{table*}

\subsubsection*{Self-consistent view of the iron line}

With the data quality presented here, it is clear that the simple prescription for the  relativistic line profile  cannot fully describe the region around the iron complex (as is made obvious by the remaining residuals in Fig.~7).

To this end, ``reflection models" \citep[i.e.][]{LightmanWhite1988, George91, Matt1991,  rossfabian1993, Zycki1994, NayakshinKazanasKallman2000, cdid2001, reflionx,  refbhb, GarciaKallman2010, GarciaKallman2011} have been produced which self-consistently   provide a treatment of the dominant atomic processes around a black hole in a variety of  ionization regimes and, given an input power-law continuum, outputs a reflection spectrum where both the ``Compton-hump", emission and absorption features are all physically linked. In this section, we proceed by replacing the \laor\ line profile with the model \reflionx\  of  \citet[][]{reflionx}. This model outputs a reflection spectrum which is calculated in the local frame, therefore we also initially employ the relativistic blurring kernel \kdblur, which is derived from the same code by \citet{laor}, to model the relativistic effects in the spectra. For the time being, we still assume a Newtonian emissivity ($q=3$).

Figure~8 shows the data-to-model ratio for this combination (Model~2 in Table~1) as well as Model~1a for ease of comparison. From Table~1 we see that replacing \laor\ with \reflionx\ indeed improves the quality of the fit ($\Delta\chisq=-79.6$ for the same number of degrees of freedom). Note that most parameters dealing with the continuum or neutral absorption have so far been very consistent between models and the only noticeable and statistically significant differences between Models~1 and 2 lie in the extent of the inner accretion disk and the normalisation of the powerlaw which is now slightly offset by the presence of the reflected continuum. Whereas the \laor\ line profile is  consistent with arising from a prograde disk around a slowly spinning black hole, the results from the more physical reflection model would imply either a disk truncated at least 1\rg\ from the ISCO of a Schwarzschild black hole or  the presence of a black hole having a retrograde spin, with an ISCO  $>6$\rg\ from the black hole.

Before we explore this possibility further, let us note that it is often the case in work employing similar methodologies to those used here, that fits with physical reflection models yield a disk radii $\sim10-20\%$ larger than that of the corresponding phenomenological fit \citep[e.g.][and references therein]{reisns,reismaxi, Waltonreis2012}.  The reason for this is that the ``iron line" complex is not limited to a single emission line and is in fact the combination of both emission and absorption edges that is not accounted for in purely emission line models such as the Gaussian profile or \laor. A further point to note is that the disk is intrinsically hot in stellar-mass black holes and for this reason any atomic features will be inherently (Compton) broadened prior to any relativistic effect that will eventually take place  \citep[see][]{refbhb}. However, the difference between the results obtained with different physical reflection models is usually fairly small in comparison to the differences between results obtained with physical reflection models and their phenomenological counterparts, particularly for sources in which high spin is inferred

\begin{figure}
\label{figure8}
\begin{center}
{\hspace*{-0.5cm}
 \rotatebox{0}{
{\includegraphics[width=8.cm]{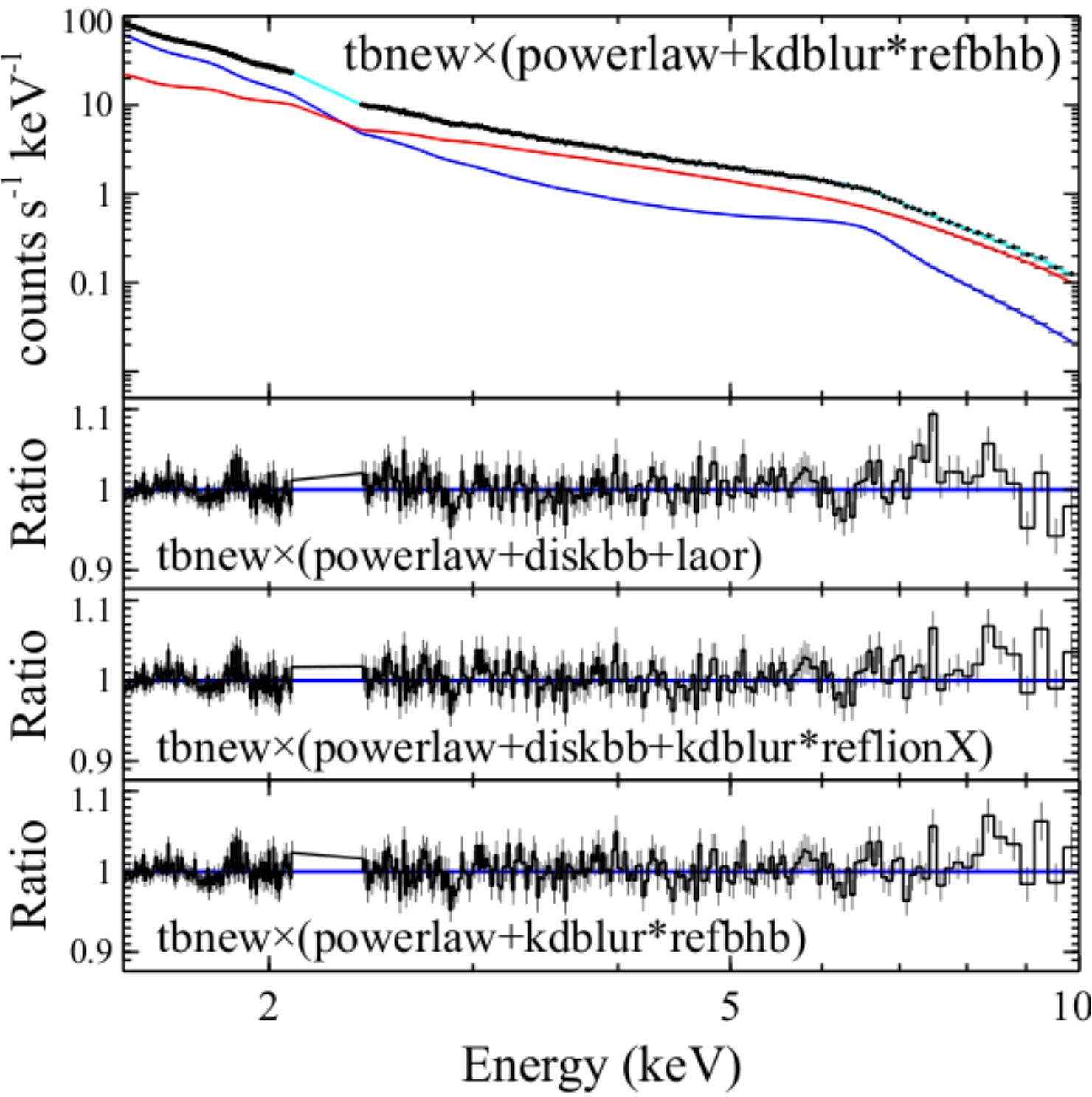} }}}
\caption{(Top:) Spectrum of \s\ fit with the self-consistent reflection model \refbhb\ together with a powerlaw (Model~3a).  The remaining panels show (from top to bottom): data/model ratio to the phenomenological Model~1a (again shown for comparison); data/model ratio to Model~2 having the phenomenological line profile replaced by the self-consistent reflection model \reflionx; and data/model ratio to the best-fit combination described as Model~3a in Table~1, where both the reflection features and the thermal disk component are modelled self-consistently with \refbhb .    }
\end{center}
\vspace{-0.5cm}
\end{figure}
                            
\begin{figure*}
\label{figure9}
\begin{center}
{\hspace*{-0.5cm}
 \rotatebox{0}{
{\includegraphics[width=13.6cm]{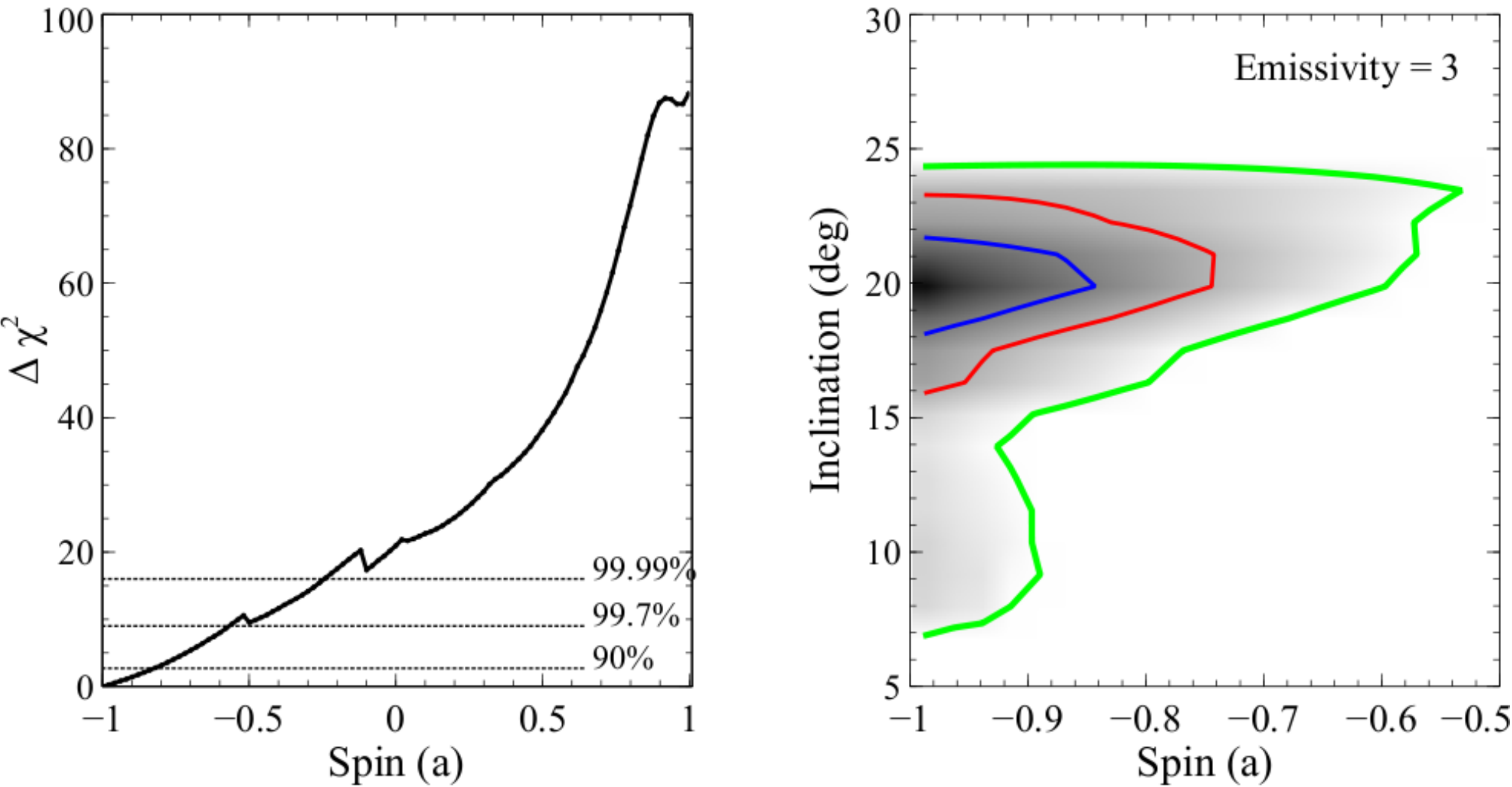} }}}
\caption{(Left:) Goodness-of-fit versus spin for Model~3b. (Right:) Contour plot investigating the possible effect of the inclination on the spin for Model~3b.  The 68, 90 and 99 per cent confidence range for two parameters of interest are shown in blue, red and green respectively. Under the assumption that the accretion disk is extending as far as the ISCO, the present result would suggest that the black hole in \s\ rapidly rotating in a retrograde manner to the accretion disk. A prograde or static Schwarzschild black hole are both rejected at greater than the $3\sigma$ level of confidence assuming a fixed emissivity index of 3. }
\vspace*{0.2cm}
\end{center}
 
  \vspace{-0.5cm}

\label{figure10}
\begin{center}
{\hspace*{-0.5cm}
 \rotatebox{0}{
{\includegraphics[width=13.6cm]{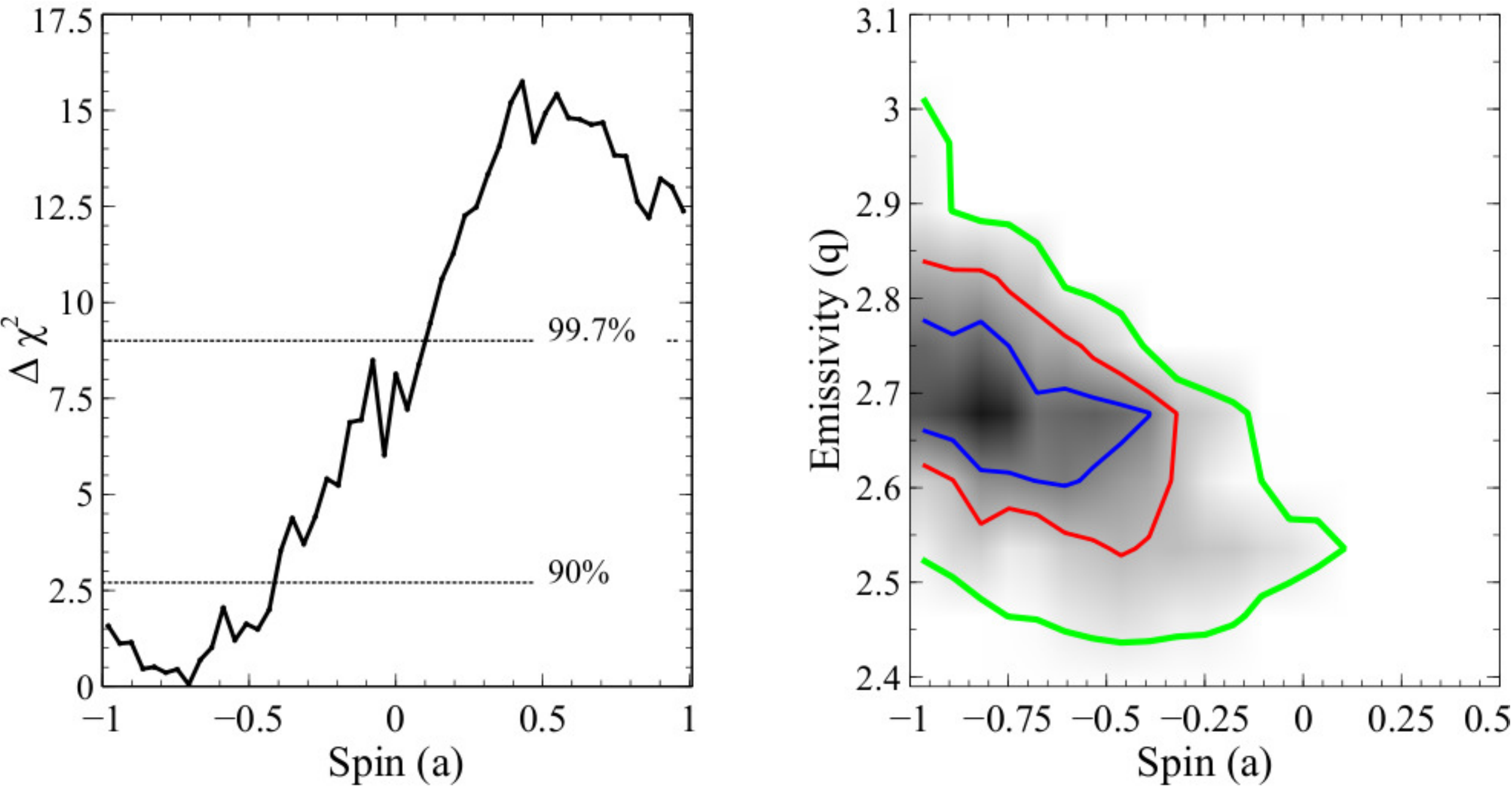} }}}
\caption{(Left:) Goodness-of-fit versus spin for Model 3b after allowing the emissivity index to vary. (Right:) Contour plot investigating the possible effect of the emissivity index on the spin for Model 3b. The 68, 90 and 99 per cent confidence range for two parameters of interest are shown in blue, red and green respectively. Despite a weaker constraint, a retrograde black hole is still preferred at the 90 per cent level of confidence.  }
\vspace*{-0.15cm}

\end{center}
\end{figure*}

With the latter point  in mind, \citet{refbhb}  developed a version of the \reflionx\ grid which is especially suited for black hole binaries. This model,  \refbhb, assumes that the   atmosphere of the accretion disk is  illuminated not only by the hard, powerlaw-like corona, but additionally by blackbody radiation intrinsic to the disk, and as such it self-consistently accounts for the disk emission, relativistic lines/edges and the reflection continuum. Detailed description of this model can be found in \citet{refbhb}  and its application to a number of black hole binaries  can be seen in \citet{reisgx, reisspin, reismaxi, Steiner2011, steinerlmcx12012, Waltonreis2012}. The parameters of the \refbhb\ model are the number density of hydrogen in the illuminated surface layer, ${\it H}_{\rm den}$, the temperature of the blackbody heating the surface layers, the power-law photon index, and the ratio of the total flux illuminating the disc to the total blackbody flux emitted by the disc. Like \reflionx, \refbhb\ outputs a  spectrum which is calculated in the local frame and we again start by  employing the convolution model \kdblur\  to model the relativistic effects in the spectra.

The above model results in an excellent fit to the data with $\chi^{2}/\nu = 1655.9/1606$ (Model~3a, see Fig.~8) which is an improvement of $\Delta\chisq=-98.6$ over the phenomenological model (Model~1b) despite it having the same number of degrees of freedom and being physically self-consistent. Note also that the \refbhb\ model accounts for both the thermal and reflection components self-consistently and as such does not require the separate addition of a \diskbb\ component as in Model~2. Although  $\xi$ is not a parameter of \refbhb, we can use the various parameters presented in Table~1 to estimate this value. The flux of the blackbody in the model is related to its temperature by the  $F_{\rm bb} \propto  T^4$ relation, therefore we can estimate $F_{bb}=(2.3\pm0.2)\times10^{21}\ergpcmsqps$.  Combining this $F_{\rm Illum}/F_{\rm BB}=0.8^{+0.1}_{-0.2}$  results in an illuminating flux of $F_{\rm Illum}= (1.8\pm0.4)\times10^{21}\ergpcmsqps$. The ionization parameter is defined as $\xi = 4\pi F/n ~\ergcmps$, is  then found by using  $n = n_H = 1.2^{+0.2}_{-0.3} \times 10^{19}{ \rm H~cm^{-3}}$ so that $\xi = 4290\pm920 ~\ergcmps$, consistent with the value found for  \reflionx\ in  Model~2.

In the present work, the $\sim 60-90\%$ increase in the inner radii between the phenomenological and self-consistent models  (see Table~1) is larger than the differences between these models typically seen from similar analyses of other black hole binaries ($\sim10-20\%$, see e.g. \citealt{reismaxi, Waltonreis2012}).  A likely cause for this larger discrepancy is that for the narrower line profile observed here, the Compton broadening intrinsically included in the \refbhb\ reflection model provides a larger relative contribution to the total width of the line than it does for cases in which high spin is inferred (i.e. sources with broader lines). The implication is that it is even more important to utilize physical reflection codes in order to reliably estimate inner disk radii when larger radii are inferred from simple phenomenological models.

\subsubsection*{A possible maximally retrograde black hole}

When using physically-motivated reflection models to reproduce the asymmetric and  moderately broad line profile,  the results point towards a accretion disk with an inner radius  beyond 6\rg\ in \s, despite the fact that the system is in a bright, active, hard intermediate state (see Figs.~1,3). In these intermediate accretion states, the disk is not expected to truncate far beyond the ISCO \citep[e.g.][and references therein]{reis20121650, Waltonreis2012}.  This is corroborated by the detection in various black hole binaries at intermediate or high accretion states, of high-frequency quasi-periodic oscillations (HFQPOs) which are often associated with Keplerian motion in the inner disk \citep[][and references therein]{MorganRemillard1997qpo, Wagoneretal2001qpo,  vanderKlisbook, Wagoner2012qpo, Belloni2012qpo}. However, due to the lack of a unique interpretation for the origin of QPOs, it is unlikely that QPOs could at this time robustly differentiate between  an accretion disk at the ISCO from that truncated at a  small radius away from the ISCO.

An intriguing possibility is that the disk is indeed extending down to the ISCO but that central back hole has a negative spin in relation to the disk angular momentum. In order to test this possibility, we replace the \kdblur\ kernel, which inherently assumes a maximally rotating, prograde black hole, with the  modern relativistic model \relconv\ of \citet[][]{relconv}, which includes the black hole spin as an additional variable parameter, allowing the dimensionless spin  to have any value from $-0.998<a<0.998$.

Initially, the emissivity index $q$ is again fixed to the Newtonian value of 3.0. This model, described in Table~1 as Model~3b, provides as equally satisfactory fit with $\chi^{2}/\nu = 1656.6/1606$ and indeed points towards the possibility of a retrograde black hole having a spin $a<-0.5$ at the $3\sigma$ level (see Fig.~9; left) if the disk is assumed to extend as far as the ISCO. In Fig.~9 (right) we also show the the 68, 90 and 95 percent confidence range for the spin as a function of the inclination of the system.  From Fig. 9, it can be seen that under the assumption that the accretion disk in \s\ is extending to the ISCO, the current data strongly imply a rapidly rotating, retrograde black hole. This conclusion remains unchanged if we allow the emissivity index to vary, as can be seen in Fig.~10, although the constraint obtained does become weaker. The results obtained in this case, referred to as Model~3c, are also presented in Table~1. A moderate improvement of $\Delta\chi^{2}=8.7$ for one additional free parameter is found, with the most notable change seen in the best fit inclination. However, other than this, the results are generally very similar to those obtained with $q=3$.

\section{discussion}

\xmm\ observed the newly discovered Galactic X-ray transient \s\  during a clear  soft-to-hard state transition (Fig.~1).  A \swift\ campaign to monitor the  outburst (\s\ is still active at the time of writing)  shows that the source at times can become  extremely soft, with well over 90\% of the 0.6-10\kev\ flux arising from a thermal disk component (Fig.~3). However, even during these times of extreme softness, at the peak of the outburst, the system is never seen to have disk temperatures much higher than $\sim0.4\kev$ (Degenaar et al; in prep.).

Observations made with the \swift\ X-ray telescope of various transient BHBs during outbursts \citep{Reynold2011swift} have shown that these transients often reach temperatures in excess of $\sim0.5$\kev\ at the peak of their outbursts. The fact that \s\ does not display such high temperatures could be a direct consequence of a retrograde system where the ISCO is at larger distances from the central black hole. However, other possibilities such as a significantly larger black hole mass cannot at this point be ruled out.  Note also that the low temperatures found for this system during outburst argues against a neutron star nature for the central object, where in this case the $\sim1$\kev\ blackbody component associated with emission from the neutron star/boundary layer would have been easily detected.

The power-spectrum shows clear band-limited noise as well as the presence of what appears to be a type-C QPO \citep{Casella2005qpo,Bellonibook2010} at approximately 4--6\hz\ (Fig.~4). The frequency of  this low-frequency QPO  and the  characteristic (break) frequency of the underlying broad-band noise components falls directly on the known  correlation  between these two parameters  (Fig.~5)  found for  black hole binaries and  low-luminosity neutron star systems (\citealt{qpofvbreak1999}; see also \citealt{Belloni2002apobreak}). The presence of a type-C QPO, coupled with an energy spectral index consistently around $\Gamma \sim 2.2$,  a 2-20\kev\ disk fraction of approximately 8\%, and the multiple soft-hard-soft transitions during this outburst (Fig.~1)  points to \s\ being a black hole binary having been caught in the hard-intermediate state as defined by \citet{Bellonibook2010}.

The clear presence of a broad, asymmetric iron emission line peaking around 7\kev\ in \s\ (Figs.~6 and 7) is indicative of reprocessed (reflected) emission  from an accretion disk relatively close to a black hole. {\color{black} During intermediate states of X-ray binaries, when   systems are relatively bright, it is generally agreed that the accretion disk is not truncated far from the central object, and in the case of black hole binaries it is likely to extend as far in as the ISCO. The presence of HFQPOs in systems such as XTE~J1550-564 during its soft-intermediate state ($\sim102-284$\hz; e.g. \citealt{HomanStates2001}; termed the VHS by the authors) or XTE~J1650-500 ($\sim50-250$\hz; \citealt{Homan2003}) during the hard-intermediate through to the soft-intermediate state, further establishes the existence of an accretion disk in the inner regions around the black hole in these states.

Indeed, spin measurements for a number of black hole binaries have been made during such intermediate states \citep{miller02j1650, hiemstra1652, reis1752, reismaxi,  Waltonreis2012}}.  For a black hole having a prograde spin, the ISCO is within 6\rg\ of the central object, with this radius decreasing as the angular momentum of the black hole increases.  Under this general picture, the results presented here  would suggest that the central black hole in \s\ is spinning in a retrograde sense, such that the ISCO is in fact at distances greater than 6\rg\ (see Table~1; Models 2 and 3). A similar claim has recently been proposed  for the AGN 3C120 \citep{Cowperthwaite2012} (although see \citet{Lohfink13}) and a retrograde black hole is also thought to be a likely explanation for A0620-00 \citep{Gou2010a0620}, where the authors find the spin to be as low as $-0.59$ at the $3\sigma$ level of confidence.

In order to test whether this hypothesis is consistent with the data and, more importantly, whether such a scenario could arise for realistic black hole masses and distances (thus setting the luminosity), we replaced the disk blackbody component  in Model~2 with the relativistic disk model \kerrbb\ \citep{kerrbb} and prevented all remaining parameters of Model~2 from varying. The spin, a parameter of \kerrbb\ was set to be -0.998, the inclination tied to that of \kdblur\ and the normalisation frozen at unity. With the remaining variable parameters being the mass-accretion rate, black hole mass and distance, we obtained a good fit with $\chi^{2}/\nu = 1665.8/1612$ and produced the distance-mass confidence contour shown in Fig.~11 for a number of different colour correction factors (\textit{f}) ranging from 1.7 to 2.5 \citep{ShimuraTakahara1995, merlonifabianross00}.  It is clear from Fig.~11 that such a retrograde black hole scenario allows for reasonable ranges in both mass and distance given a reasonable range in colour correction factor.

If \s\ indeed harbours a retrograde black hole, then we must question how such a system came about. The real  possibility of retrograde, stellar mass black holes is not new \citep[see e.g.][]{ShapiroLightman1976, FryxellTaam1988flip}, and reversals  in the rotation direction of an accretion disk have at times been invoked to explain state transitions in \cyg\ \citep{ShapiroLightman1976, Zhangetal19971655qpo}. However, for such a scenario to take place, accretion must happen with random density perturbations in the accreting gas or/and asymmetries so that these perturbations in angular momentum can compete with or dominate the net angular momentum of the disk and cause ``flip-flop" instability \citep[e.g.][and references therein]{ShapiroLightman1976,Matsuda1987flip,FryxellTaam1988flip, BlondinPope2009flip}. Such a scenario could be produced from a fluctuating stellar wind originating in a massive secondary in systems like \cyg\ but is not immediately clear whether similar processes could take place around low-mass X-ray binaries such as \s. 

Another intriguing possibility, is that the compact binary system was formed via the tidal capture of a secondary star \citep{Fabian1975tidal}, possibly in a globular cluster environment, at which point the retrograde nature of the system could have been locked at the time of the capture. The nearest globular cluster, NGC 6712 is approximately 630\pc\ away from \s\  \footnote{Assuming \s\ is at a distance of 6.9\kpc\ from the sun, similar to NGC 6712} and assuming a typical natal kick velocity of a few  hundred \kmps\ \citep[e.g.][]{Fabian1975tidal,Mirabelkick2001, gonzaleskick2006}, it would have taken the system a few million years to arrive at its current location, only a small fraction of its lifetime. Furthermore, the alignment timescale  in such systems is usually a significant fraction of their lifetime \citep[e.g.][]{Maccarone2002misalignment, MartinToutPringle2008j1655, MartinReisPringle2008}, meaning that once the system is formed in a retrograde manner, it could maintain this configuration  over most of its lifetime.

\begin{figure}[!t]
\label{figure11}
\begin{center}
{\hspace*{-0.5cm}
 \rotatebox{0}{
{\includegraphics[width=8.2cm]{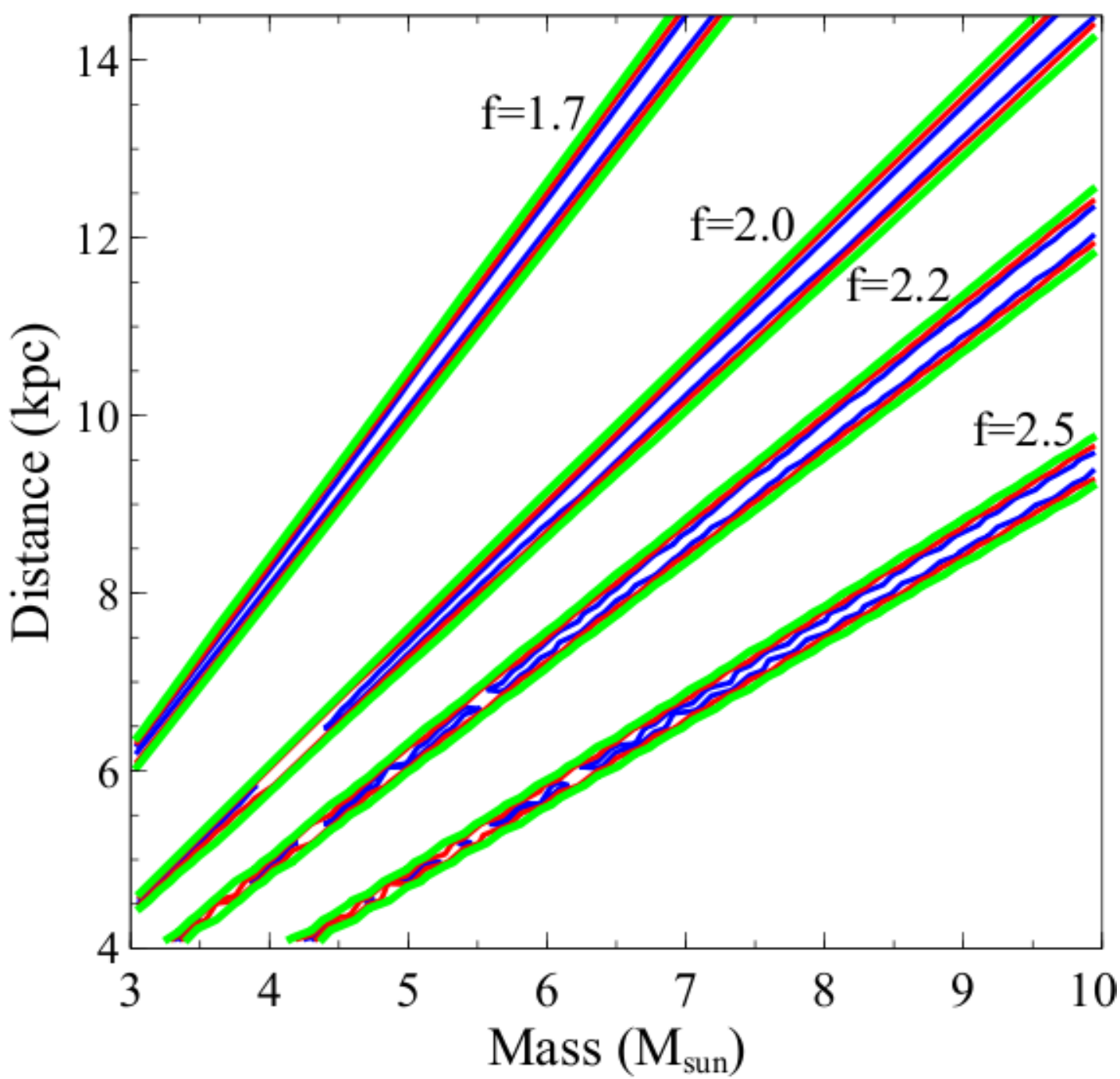} }}}
\caption{ Contour plots investigating  possible range in mass and distances which would allow \s\ to be a maximally retrograde black hole at the observed flux level.  The 68, 90 and 95 per cent confidence range for two parameters of interest are shown in black, red and green respectively for a number of different colour correction factors (\textit{f}; see text).  }
\end{center}
\end{figure}

However, there are also a number of possible scenarios that do not require the presence of a retrograde black hole. One such possibility is that the disk is undergoing a transition to a radiatively inefficiently accretion flow (RIAF) at $r\sim9$\rg\ thus causing the inner regions to be optically thin and fully ionised, so limiting the iron line emission and other reflection signatures to regions beyond this radius. In principle, the iron line profile arising from an inner radius of  9~\rg\ is different between a scenario where the disk is truncated around a prograde black hole to that expected from a disk a the ISCO of a maximally retrograde black hole (e.g. \citealt{relconv}),  however, the expected differences are too subtle to be observed  with the current data. Indeed, modifying Model~3b so that the black hole is assumed to have a maximal prograde spin, we obtain an inner disk radius of  $r_{\rm in}=9.5^{+2.6}_{-1.1}$\rg, and a goodness of fit similar to above.

Such radiatively inefficiently accretion flows are indeed observed in stellar mass black hole binaries. However they are often observed only at much lower fluxes, at luminosities $ \lesssim1.5\times 10^{-3} L_{Edd}$ (\citealt{reislhs, Reynold2011swift}; see also \citealt{Rykoff2007J1817, tomsick09gx, Reynolds2010lhs, reisj1118,reis20121650, Waltonreis2012}) as opposed to the observed intermediate state with a luminosity of  $\sim 5\%$ Eddington. Nonetheless, it is still possible that we are observing \s\ during the onset of the RIAF  state when the system is still accreting at a higher rate. One potential explanation for a truncated disk at such a high Eddington rate would be the possibility  that the inner region is undergoing continuous evaporation and re-condensation as the outburst progresses, and our observation caught the source at a time when the inner region was fully evaporated. Such ``disk evaporation" models for both AGN and BHBs are discussed in detail in \citet{TaamLiu2012, LiuTaam2009, Taametal2008} and references therein.

{\color{black} Lastly, it is also plausible that the inner regions may have been evacuated by a jet in a similar manner to that proposed by \citet{Cowperthwaite2012, Lohfink13}  for  the AGN 3C120.  \s\ was observed in radio at the peak of the outburst (\citealt{ATel4295}; see Fig.~1) approximately two months before the current observation.  Given that the viscous time scale for the accretion disk at radius $r$ is $t_{visc}(r) = \frac{2}{3\alpha}(\frac{h}{r})^{-2}\frac{1}{\Omega_{K}(r)}$-s \citep{shakuraSunyaev73}, where $\alpha$ is the viscosity parameter, $h$ describes the disk scale height  where for a geometrically-thin disk $h/r\ll 1$, and $\Omega_{K}(r)$ is the Keplerian frequency of the disk at radius $r$, we would expect any potential ``hole" in the inner $\sim10$\rg\ to be filled on time scales of the order of 100s seconds.\footnote{Assuming $M_{BH}=8\msun$, $h/r=0.01$ and $\alpha = 0.1$ results in a viscous time of $\sim 83$~seconds at 10\rg.} Unless \s\ is continuously ejecting its inner disk in the form of a (radio) jet, the short viscous time scale expected for BHB systems suggest that the iron line observed from \s\ is indeed tracing the ISCO of a retrograde black hole. In order to validate this hypothesis we would need simultaneous radio-X-ray observations so as to establish a casual connection between a possible ``empty" inner accretion disk and continuous jet emission. }

As a final note, we stress that, even assuming the disk does extend to the ISCO,
the requirement for a retrograde spin in \s\ is only significant at the $3\sigma$ level of confidence (see Fig.~10), and there is therefore a non-negligible chance that we are simply witnessing a black hole with near-zero angular momentum. If this is the case, the observed radio flare (Fig.~1) would suggest that black holes with substantial angular momentum are not necessary for such events to occur, and shed further insight into the possible role of black hole spin in launching radio jets \citep[][]{fender2010jets,NarayanMcClintock2012, Waltonreisspin2013, AKing13}.

\vspace{0.1cm}
\section{Acknowledgements}
RCR thanks the Michigan Society of Fellows and NASA. A
further thank you goes to the \xmm\ team, which scheduled
this TOO observation, and to the \swift\ team for providing
prompt notice of this new source to the astronomical community
at large. RCR is supported by NASA through the Einstein
Fellowship Program, grant number PF1-120087 and is a
member of the Michigan Society of Fellows.
\bibliographystyle{mnras}
\bibliography{/Users/rdosreis/papers/bibtex.bib}

\end{document}